# RECIPROCITY INDEPENDENT LORENTZ TRANSFORMATION

Mushfiq Ahmad
Department of Physics, Rajshahi University, Rajshahi, Bangladesh
e-mail: mushfiqahmad@ru.ac.bd

## Abstract

We have defined slowness (or reciprocal velocity, corresponding to velocity *v*) as *cc/v,* where *c* is the speed of light. It is observed that the relative velocity remains invariant if the velocities are replaced by corresponding slownesses i.e. relative motion in one dimension is reciprocal symmetric. Reciprocity operation, which converts a velocity to the corresponding slowness, is found. Lorentz transformation is generalized so that Lorentz invariance is maintained if velocities are replaced by corresponding slownesses.

## 1. Introduction

Consider a vector in a 2 dimensional Cartesian space, **V**= x**i** + y**j**. Interchanging between x and y, we get the vector **V'**=y**i** + x**j**. **V'** is rotated with respect to **V,** but the length remains unchanged, |**V'**|=|**V**|. We now consider relativistic motion in 1 space dimension. It is an event in a 2 dimensional Space-Time space. Interchanging between space and time components should only rotate it in the 2 dimensional space keeping the Lorentz length invariant (Lorentz invariant). But interchanging between space and time components means changing velocity to its reciprocal. Therefore, reciprocation should maintain Lorentz invariance. This we shall study below.

## 2. Motion in One Space Dimension

We shall measure velocity, *v*, in units of *c*. Therefore, our velocity will be *v/c*. We shall define the corresponding slowness or reciprocal velocity, $v^*$, by the relation

$$v^*.v = c^2 \tag{2.1}$$

If *u* is the velocity of a moving body and *v* is the velocity of the observer, the relative velocity is

$$u \oplus (\pm v) = \frac{u \pm v}{1 \pm \frac{u.v}{c^2}} \tag{2.2}$$

We observe that the relative velocity remains invariant if the velocities are replace by corresponding slownesses. This we shall call reciprocal symmetry.[1] Using (2.1)

$$u \oplus (\pm v) = (u^*) \oplus (\pm v^*) \tag{2.3}$$

*x* is a distance covered in time *t*, as observed by a an observer at rest. The distance *x'* and time *t'* as observed by an observer moving with velocity *v* are[2]

$$x' = \frac{x - vt}{\sqrt{1 - (v/c)^2}} \tag{2.4}$$

$$t' = \frac{t - vx/c^2}{\sqrt{1-(v/c)^2}} \tag{2.5}$$

Lorentz invariance requirement is

$$(ct')^2 - (x')^2 = (ct)^2 - x^2 \tag{2.6}$$

## 3. Rotation in Reciprocity Space

Let $\tilde{v}$ and $\tilde{x}$ be $v$ and $x$ rotated in reciprocity space through angle $\varphi$

$$\tilde{v} = \frac{v + ic\tan(\phi/2)}{1 + i(v/c)\tan(\phi/2)} = \frac{v + icr}{1 + i(v/c)r} \tag{3.1}$$

And

$$\tilde{x} = \frac{x + ict\tan(\phi/2)}{1 + i(x/ct)\tan(\phi/2)} = \frac{v + ictr}{1 + i(x/ct)r} \tag{3.2}$$

where

$$\tan(\phi/2) = r \tag{3.3}$$

(3.1) gives the reciprocal and agrees with (2.1) when $\varphi = \pi$.

$$\tilde{v} \xrightarrow[\phi\to\pi]{} v^* \tag{3.4}$$

We define the reciprocal of $x$ corresponding to (2.1) by

$$x^*.x = (ct)^2 \tag{3.5}$$

(3.2) gives the reciprocal and agrees with (3.5) when $\varphi = \pi$.

$$\tilde{x} \xrightarrow[\phi\to\pi]{} x^* \tag{3.6}$$

## 4. Reciprocity Independent Lorentz Transformation

Let reciprocity independent Lorentz transforms of $x$ and $t$ be

$$\tilde{x}' = g\{(\tilde{x} - \tilde{v}t)\} \tag{4.1}$$

$$\tilde{t}' = g\left(t - \tilde{x}\tilde{v}/c^2\right) \tag{4.2}$$

where

$$g = \frac{\{1 + i(x/ct)r\}\{1 + i(v/c)r\}}{(1+r^2)\sqrt{1-(v/c)^2}} \tag{4.3}$$

Transformations (4.1) and (4.2) ensure Lorentz invariance for all reciprocity states (values of $\varphi$).

$$(c\tilde{t}')^2 - \tilde{x}'^2 = (ct)^2 - x^2 \tag{4.4}$$

## 5. Motion in Three Space Dimensions

We define reciprocals $\mathbf{V}^*$ and $\mathbf{X}^*$ of $\mathbf{V}$ and $\mathbf{X}$ by the relations (see (2.1) and (3.5))

$$\mathbf{V}^*.\mathbf{V} = c^2 \;\; and \;\; \mathbf{X}^*.\mathbf{X} = (ct)^2 \tag{5.1}$$

We define $\widetilde{\mathbf{V}}$ and $\widetilde{\mathbf{X}}$, $\mathbf{V}$ and $\mathbf{X}$ rotated in reciprocity space through $\varphi$, by
We shall choose the definitions with arbitrary $\mathbf{r}$

$$\mathbf{V}^* = c^2 \frac{\left(1-\sqrt{1-(\mathbf{V}/c)^2}\right)\frac{\mathbf{r.V}}{\mathbf{V}^2}\mathbf{V} + \mathbf{r}\sqrt{1-(\mathbf{V}/c)^2}}{\mathbf{r.V}} \tag{5.2}$$

And

$$\mathbf{X}^* = (ct)^2 \frac{\left(\mathbf{1}-\sqrt{\mathbf{1}-(\mathbf{X}/ct)^2}\right)\frac{\mathbf{r.X}}{\mathbf{X^2}}\mathbf{X} + \mathbf{r}\sqrt{\mathbf{1}-(\mathbf{X}/ct)^2}}{\mathbf{r.X}} \tag{5.3}$$

The above definitions fulfill requirements (5.1).

Corresponding to (3.1) and (3.2) we define rotated $\mathbf{V}$ and $\mathbf{X}$ by

$$\widetilde{\mathbf{V}} = \frac{\mathbf{V} + \left(1-\sqrt{1-(\mathbf{V}/c)^2}\right)\frac{ic\mathbf{r.V}}{\mathbf{V}^2}\mathbf{V} + ic\mathbf{r}\sqrt{1-(\mathbf{V}/c)^2}}{\mathbf{1}+\frac{ir\mathbf{.V}}{c}} \tag{5.4}$$

And

$$\widetilde{\mathbf{X}} = \frac{\mathbf{X} + \left(\mathbf{1}-\sqrt{\mathbf{1}-(\mathbf{X}/ct)^2}\right)\frac{ict\mathbf{r.X}}{\mathbf{X^2}}\mathbf{X} + ict\mathbf{r}\sqrt{\mathbf{1}-(\mathbf{X}/ct)^2}}{\mathbf{1}+\frac{i\mathbf{r.X}}{ct}} \tag{5.5}$$

Where

$$\mathbf{r} = \tan(\varphi/2)\mathbf{n} \; with \; |\mathbf{n}| = 1 \tag{5.6}$$

$$\widetilde{\mathbf{V}} \xrightarrow[\phi\to\pi]{} \mathbf{V}^* \tag{5.7}$$

And

$$\widetilde{\mathbf{X}} \xrightarrow[\phi\to\pi]{} \mathbf{X}^* \tag{5.8}$$

We now define Lorentz transforms of $\widetilde{\mathbf{X}}$ and $t$ by

$$\widetilde{\mathbf{X}}' = G\left\{\left(\sqrt{1-(\mathbf{V}/c)^2}\right)\widetilde{\mathbf{X}} + \left[\left(1-\sqrt{1-(\mathbf{V}/c)^2}\right)\frac{\widetilde{\mathbf{X}}.\widetilde{\mathbf{V}}}{\widetilde{\mathbf{V}}^2} - t\right]\widetilde{\mathbf{V}}\right\} \tag{5.9}$$

and

$$\widetilde{t}' = G\left\{t - \frac{\widetilde{\mathbf{X}}.\widetilde{\mathbf{V}}}{c^2}\right\} \tag{5.10}$$

where

$$G = \frac{\{c + i\tan(\phi/2)\,\mathbf{X.n}/t\}\,\{c + i\tan(\phi/2)\,\mathbf{V.n}\}}{c^2\{1+[\tan(\phi/2)]^2\}\sqrt{1-(\mathbf{V}/c)^2}} \tag{5.11}$$

Lorentz invariance relation is

$$(c\widetilde{t}')^2 - (\widetilde{\mathbf{X}}')^2 = (ct)^2 - \mathbf{X}^2 \tag{5.12}$$

## 8. Conclusion

We have been able to generalize **V** and **X** to cover all reciprocity states. We have also found their Lorentz transforms, which maintain Lorentz invariance for all reciprocity states.

---